\documentclass[11pt,a4paper,english,amssymb,nofootinbib,superscriptaddress,showpacs]{revtex4}
\usepackage{graphicx,amscd,amsmath,amssymb,verbatim}
\usepackage{lmodern}
\usepackage{esint}

\makeatletter
\usepackage{lmodern}\usepackage{babel}

\pdfpageheight\paperheight
\pdfpagewidth\paperwidth

\usepackage{latexsym}\usepackage{bm}

\usepackage[colorlinks,hyperindex, % pagebackref,
            bookmarks=true,bookmarksopen=true]{hyperref}
    \hypersetup
    {
        colorlinks=true,
        linkcolor=blue,
        urlcolor=blue,
        filecolor=black,
        citecolor=red,
        pdfstartview=FitV,
        pdftitle={},
        pdfauthor={Ilmar Gahramanov, Elmar Asgerov},
        pdfsubject={},
        pdfkeywords={},
        pdfpagemode=None,
        bookmarksopen=true
    }

\usepackage[notcite, color, final]{showkeys}
\definecolor{refkey}{gray}{.5}
\definecolor{labelkey}{gray}{.5}

\newcommand\Tr[1]{\mbox{tr}\left(#1\right)}

\begin{document}
\title{A remark on the trace-map for the Silver mean sequence}

\author{Ilmar Gahramanov\footnote{Currently the author is located at DESY Theory Group, Germany, \href{mailto:ilmar.gahramanov@desy.de}{ilmar.gahramanov@desy.de}}}
\affiliation{Department of Theoretical Physics and Quantum Technologies,
National University of Science and Technology ``MISIS'', Moscow, Russia}
\author{Elmar Asgerov}
\affiliation{Institute of Radiation Problems ANAS, Baku, Azerbaijan}
\affiliation{Frank Laboratory of Neutron Physics, JINR, Dubna, Russia}

\date{\today}

\begin{abstract}
In this work we study the Silver mean sequence based on substitution rules by means of a transfer-matrix approach.
Using transfer-matrix method we find a recurrence relation for the traces of general transfer-matrices which characterizes electronic properties of the quasicrystal in question. We also find an invariant of the trace-map.
\end{abstract}
\maketitle

\section{Introduction}

Since the discovery of quasicrystals in nature \cite{Shech}, aperiodic structures have been extensively studied in condensed matter physics (see for e.g. \cite{Vek}--\cite{Macia3}) and also in mathematics (see \cite{Quasicrys, DirMath}). One of the main questions in the theory of one-dimensional aperiodic structures is to find the relationship between their atomic topological order and the physical properties \cite{Macia2}, \cite{Alb}. The 1D aperiodic sequences are characterized by the nature of their Fourier spectrum. We define a 1D quasicrystal as a 1D aperiodic sequence with pure point Fourier spectrum. In this article we concentrate on the study of the Silver mean sequence, which is common 1d quasicrystal.

A convenient way to study the electronic spectrum of a 1D quasicrystals is the trace-map technique, introduced by Kohmoto, Kadanoff and Tang \cite{KKT}. Using this method the relations for transfer-matrices and their traces for the Fibonacci sequence were obtained \cite{KKT, Koh}. For generalizations of the trace-map to other one-dimensional aperiodic sequences see \cite{Cheng}--\cite{GumAli2} and references in \cite{Macia2}.

%\vspace{0.2cm}

1D quasicrystals can be described by the 1D discrete Schr\"{o}dinger equation with quasiperiodic potential
\begin{equation}
\label{Sch} \psi_{n+1}+\psi_{n-1}+\epsilon_n\psi_n=E\psi_n,
\end{equation}
where $\psi_n$ is a wave function on $n$th site, $\epsilon_n$ could take two values $\epsilon_A$ or $\epsilon_B$. The equation (\ref{Sch}) has been studied in many articles (see, for example, references in \cite{Macia2, Alb}) and it is an adequate model for considering qualitatively the effects of the aperiodicity on the electronic structure.
One can rewrite the Schr\"{o}dinger equation (\ref{Sch}) in the matrix form
\begin{equation}
\left(
\begin{gathered}
  \psi _{n + 1}  \hfill \\
  \psi _n  \hfill \\
\end{gathered}  \right) = M(n) \left( \begin{gathered}
  \psi _n  \hfill \\
  \psi _{n - 1}  \hfill \\
\end{gathered}  \right),
\end{equation}
where
\begin{equation}
M(n)  = \left( {\begin{array}{*{20}c}
   {E - \epsilon_n} & { - 1}  \\
   1 & 0  \\
 \end{array} } \right)
\end{equation}
is the so-called transfer-matrix at the $n$'th site. The general transfer-matrix $M_n$ connecting the wave function
$(\psi_{n+1}, \psi_{n})$ and $(\psi_1,\psi_0)$  defined by
\begin{equation}
\left(
\begin{gathered}
  \psi _{n + 1}  \hfill \\
  \psi _n  \hfill \\
\end{gathered}  \right) = M_n \left( \begin{gathered}
  \psi _1  \hfill \\
  \psi _0  \hfill \\
\end{gathered}  \right),
\end{equation}
with the general transfer-matrix
\begin{equation}
M_n=M(n)\cdot M(n-1)\ldots M(1).
\end{equation}

Solving the Schr\"{o}dinger equation (\ref{Sch}) is equivalent to calculate products of transfer-matrices. The allowed regions of the energy spectrum are determined by the condition
\begin{equation}
|\Tr{M_n}|=|\prod \Tr{M(n)}|\leq 2.
\end{equation}
From a mathematical point of view this condition means that is that if $M(n)$ is the matrix of an area-preserving linear transformation of a plane into itself, then the mapping is stable if $|\Tr{M_n}|< 2$, and unstable if $|\Tr{M_n}|> 2$ (for more details see \cite{Arn}).

\section{Silver mean sequence}

To construct the Silver mean sequence we use the two-letter substitution rules:
\begin{equation}
A \rightarrow B, ~~~~ B \rightarrow BBA.
\end{equation}
One can furthermore construct the Silver mean sequence in analogue to the Pell numbers \cite{Pell}
\begin{equation}
P_n=2P_{n-1}+P_{n-2}
\end{equation}
with $P_1=A$ and $P_2=B$. Some authors call this sequence `` intergrowth sequence'' \cite{Huang1}--\cite{Huang3} or `` octonacci sequence'' \cite{Mos}--\cite{Grimm}.

The corresponding general transfer-matrix $M_n$ for the Silver mean sequence can be written as follows:
\begin{equation}
\label{main} M_{n}=M_{n-1}^2 M_{n-2}.
\end{equation}
This expression can be proven by induction \cite{Bom}.

Using the Cayley-Hamilton theorem for $2\times 2$ matrices $M_n$ with $det M_n=1$, i.e.
\begin{equation}
M_n^{-1}+M_n=\Tr{M_n},
\end{equation}
one can easily obtain the following expression:
%\begin{widetext}
\begin{eqnarray} \nonumber
\label{0} M_{n+1} &=& M_{n-1}(\Tr{M_n})^2-M_{n-1}M_n^{-1}-M_{n-1}\\ \nonumber
                  &=& M_{n-1}(\Tr{M_n})^2-M_{n-1}M_n^{-1}\frac {\Tr{M_{n-1}}} {\Tr{M_{n-1}}}-M_{n-1}\\ \nonumber
									&=& M_{n-1}(\Tr{M_n})^2-\frac{M_{n-1}^2M_n^{-1}+M_n^{-1}}{\Tr{M_{n-1}}}-M_{n-1}.
\end{eqnarray}
By taking the trace of (\ref{0}) and using the expression $\Tr{M_{n-2}}=\Tr{M_{n-1}^2M_n^{-1}}$ and $\Tr{M}=\Tr{M^{-1}}$ we find the recurrence relation for the traces of the general transfer-matrices
\begin{equation}
\label{main} \boxed{\Tr{M_{n+1}}=\Tr{M_{n-1}}(\Tr{M_n})^2-\frac {(\Tr{M_n})^2+\Tr{M_{n-2}}\Tr{M_n}} {\Tr{M_{n-1}}}-\Tr{{M_{n-1}}}.}
\end{equation}
In contrast to \cite{GumAli, GumAli2} we show that the recurrence relation (\ref{main}) can be expressed in terms of $\Tr{M_i}$'s only. This recurrence relation is useful for analytical calculations, as well for numerical computations. From the physical point of view these traces are important because they determine the structure of the energy spectrum of quasiperiodic sequence. For instance, using (\ref{main}) one can calculate forbidden and allowed regions in the energy spectrum \cite{GumAli}, the Lyapunov exponent \cite{Luck}, etc. For other physical applications see \cite{Macia, Macia2, Alb, Luck} and also \cite{Lu, Roche}.
%\end{widetext}

Following \cite{Koh} one can define a three-dimensional vector $r=(x,y,z)$, where $x=\frac{1}{2}\Tr{M_{n+1}}$, $y=\frac{1}{2}\Tr{M_n}$, $z=\frac{1}{2}\Tr{M_{n-1}}$ and then alternatively express (\ref{main}) as
\begin{equation}
r_{n+1}=f(r_{n}).
\end{equation}
This nonlinear map has an invariant, i.e. it is the same for any $n$'th generation:
\begin{equation}
\label{inv} \boxed{I=-xz+\left(\frac{x+z} {2y}\right)^2+y^2-1.}
\end{equation}
If we now redefine $z$ as
\begin{equation}
z\rightarrow 4y^2x-x-4yz,
\end{equation}
expression (\ref{inv}) becomes
\begin{equation}
\label{inv2} I=x^2+y^2+4z^2-4xyz-1.
\end{equation}
This is the same result which was found in \cite{GumAli, GumAli2}.

\section{Discussion}

In this paper we considered the trace-map associated with the Silver mean sequence. We found the recurrence relation for the trace of the general transfer-matrix of the Silver mean sequence. We have shown that this recurrence relation can be expressed in terms of $\Tr{M_i}$'s only. Finally we found an invariant of the trace-map. The recurrence relation and invariant of the trace-map are closely related to the spectral properties of the sequence in question\footnote{For physical mean of the invariant see \cite{KKT}, \cite{Koh} and also the recent preprint \cite{Chak}.}. This invariant plays an important role in understanding quasiperiodic sequences \cite{Iguchi}.

\vspace{0.3cm}

{\bf Acknowledgments.} I.G. would like to especially thank Edvard Musaev, Boris Kheyfets and David Klein for interesting discussions and for their help in the preparation of the manuscript. The authors are grateful to Gulmammad Mammadov for useful comments.


\begin{thebibliography}{40}

\bibitem{Shech} D.~Shechtman, I.~Blech, D.~Gratias, J.~W.~Cahn, \textit{Metallic Phase with Long-Range Orientational Order and No Translational Symmetry}, \href{http://prl.aps.org/abstract/PRL/v53/i20/p1951_1}{Phys. Rev. Lett. \textbf{53}, 1951 (1984)}.

\bibitem{Vek} Y.~Kh.~Vekilov and M.~A.~Chernikov, \textit{Quasicrystals}, \href{http://iopscience.iop.org/1063-7869/53/6/R01}{Phys.-Usp. \textbf{53} 537 (2010)}.

\bibitem{Macia} E.~Macia, \textit{The role of aperiodic order in science and technology}, \href{http://iopscience.iop.org/0034-4885/69/2/R03}{Rep. Prog. Phys. \textbf{69} 397 (2006)}.

\bibitem{Macia2} B.~E.~Macia, \textit{Aperiodic structures in condensed matter : fundamentals and applications}, Taylor \& Francis (2008).

\bibitem{Macia3} E.~Macia, \textit{Exploiting aperiodic designs in nanophotonic devices},  \href{http://iopscience.iop.org/0034-4885/75/3/036502}{Rep. Prog. Phys. 75 036502 (2012)}.

\bibitem{Quasicrys} \textit{Quasicrystals -- An Introduction to Structure, Physical Properties and Applications}, edited by J.-B.~Suck, M.~Schreiber and P.~ H\"aussler, Springer, Berlin, (2002).

\bibitem{DirMath} \textit{Directions in Mathematical Quasicrystals}, edited by M.~Baake \& R.~Moody, CRM Monograph Series, Vol. 13, American Mathematical Society (2000).

\bibitem{Alb} E.~L.~Albuquerque and M.~G.~Cottam, \textit{Theory of elementary excitations in quasiperiodic structures}, \href{http://www.sciencedirect.com/science/article/pii/S0370157302005598}{Phys. Reports \textbf{376}, 225-337 (2003)}.

%\bibitem{Baake0} M. Baake, A guide to mathematical quasicrystals, \href{http://arxiv.org/abs/math-ph/9901014}{arXiv:math-ph/9901014v1}

\bibitem{KKT} M.~Kohmoto, L.~P.~Kadanoff, and C.~Tang, \textit{Localization Problem in One Dimension: Mapping and Escape}, \href{http://prl.aps.org/abstract/PRL/v50/i23/p1870_1}{Phys. Rev. Lett. \textbf{50}, 1870 (1983)}.

\bibitem{Koh} M.~Kohmoto, B.~Sutherland, and C.~Tang, \textit{Critical wave functions and a Cantor-set spectrum of a one-dimensional quasicristal model},  \href{http://prb.aps.org/abstract/PRB/v35/i3/p1020_1}{Phys. Rev. \textbf{B 35}, 1020 (1987)}.

\bibitem{Cheng} S.~Cheng and G.~Jin, \textit{Trace map and eigenstates of a Thue-Morse chain in a general model}, \href{http://prb.aps.org/abstract/PRB/v65/i13/e134206}{Phys. Rev. \textbf{B 65}, 134206 (2002)}.

\bibitem{Kolar} M.~Kolar, M.~K.~Ali, F.~Nori, \textit{Generalized Thue-Morse chains and their physical properties} \href{http://prb.aps.org/abstract/PRB/v43/i1/p1034_1}{Phys. Rev. \textbf{B 43}, 1034 (1991)}.

\bibitem{Kolar2} M.~Kolar, M.~K.~Ali, Trace maps associated with general two-letter substitution rules, \href{http://pra.aps.org/abstract/PRA/v42/i12/p7112_1}{Phys. Rev. \textbf{A 42}, 7112 (1990)}.

\bibitem{Ghosh} A.~Ghosh and S.~N.~Karmakar, \textit{Trace map of a general aperiodic Thue-Morse chain: Electronic properties}, \href{http://prb.aps.org/abstract/PRB/v58/i5/p2586_1}{Phys. Rev. \textbf{B 58}, 2586 (1998)}.

\bibitem{Kar} S.~Sil, S.~N.~Karmakar, R.~K.~Moitra, A.~Chakrabarti, \textit{Extended states in one-dimensional lattices: Application to the quasiperiodic copper-mean chain}, \href{http://prb.aps.org/abstract/PRB/v48/i6/p4192_1}{Phys. Rev. \textbf{B 48}, 4192 (1993)}.

\bibitem{GumAli} G.~Gumbs, M.~K.~Ali, \textit{Electronic properties of the tight-binding Fibonacci Hamiltonian}, \href{http://iopscience.iop.org/0305-4470/22/8/012}{J.Phys. A: Math. Gen. \textbf{22}, 951-970 (1989)}.

\bibitem{GumAli2} G.~Gumbs, M.~K.~Ali, \textit{Dynamical Maps, Cantor Spectra, and Localization for Fibonacci and Related Quasiperiodic Lattices}, \href{http://prl.aps.org/abstract/PRL/v60/i11/p1081_1}{Phys.Rev.Lett. \textbf{60}, 1081 (1988)}.

\bibitem{Arn} \textit{Ordinary Differential Equations} by Vladimir I. Arnol'd, Springer-Verlag: Berlin, Heidelberg (1992).

\bibitem{Pell} M.~Bicknell, \textit{A primer on the Pell sequence and related sequences}. Fibonacci Quart. \textbf{13}, 4, 345--349 (1975); \href{http://www.fq.math.ca/Scanned/13-4/bicknell.pdf}{MR0387173}.

\bibitem{Huang1} X.~Huang and Y.~Liu, \textit{Spectral Structure and Gap-Labeling Properties for a New Class of One-Dimensional Quasilattices}, \href{http://iopscience.iop.org/0256-307X/9/11/012/}{Chinese Phys. Lett. \textbf{9} 609 (1992)}.

\bibitem{Huang2} X.~Q.~Huang, Y.~Y.~Liu, D.~Mo, \textit{Electronic properties of one-dimensional quasilattices} \href{http://www.springerlink.com/content/w21h322776216354/}{Z. Phys. B 93, 103 (1993)}.

\bibitem{Huang3} Y.~Wang, X.~Q.~Huang, C.~Gong, \textit{Light Transmission Through Symmetric Fibonacci-Class Multilayers}, \href{http://iopscience.iop.org/0256-307X/17/7/012}{Chinese Phys. Lett. \textbf{17} 498 (2000)}.

\bibitem{Mos} S.~R.~Mosseri, and J.--F.~Sadoc, \textit{Geometric study of a 2D tiling related to the octagonal
quasiperiodic tiling}, \href{http://jphys.journaldephysique.org/index.php?option=com_article&access=doi&doi=10.1051/jphys:0198900500240346300&Itemid=129}{J. Phys. France, \textbf{50}, 3463-3476 (1989)}.

\bibitem{Crys} \textit{Crystallography of Quasicrystals: Concepts, Methods and Structures}, by Walter Steurer and Sofia Deloudi, Springer-Verlag: Berlin, Heidelberg (2009).

\bibitem{Grimm} H.~Q.~Yuan, U.~Grimm, P.~Repetowicz, and M.~Schreiber, \textit{Energy spectra, wave functions, and quantum diffusion for quasiperiodic systems}, \href{http://prb.aps.org/abstract/PRB/v62/i23/p15569_1}{Phys. Rev. \textbf{B62}, 15569 (2000)}.

\bibitem{Bom} E.~Bombieri, J.~E.~Taylor \textit{Which distributions of matter diffract? An initial investigation}, J. Physique Coll. \textbf{C3} 19--29 (1986).

\bibitem{Luck} J.~M.~Luck,  \textit{Cantor spectra and scaling of gap widths in deterministic aperiodic systems}, \href{http://prb.aps.org/abstract/PRB/v39/i9/p5834_1}{Phys. Rev. \textbf{B 39}, 9, 5834 (1989)}.

\bibitem{Lu} J.~P.~Lu, T.~Odagaki, J.~L.~Birman, \textit{Properties of one-dimensional quasilattices}, \href{http://prb.aps.org/abstract/PRB/v33/i7/p4809_1}{Phys. Rev. \textbf{B 33}, 4809 (1986)}.

\bibitem{Roche} S.~Roche, G.~Trambly de Laissardiere, and D.~Mayou, \textit{Electronic transport properties of quasicrystals}, \href{http://jmp.aip.org/resource/1/jmapaq/v38/i4/p1794_s1}{J. Math. Phys. 38, 1794 (1997)}.

\bibitem{Chak} A.~Chakrabarti, S.~Chattopadhyay, \textit{A Fibonacci atomic chain with side coupled quantum dots: crossover from a singular continuous to a continuous spectrum and related issues}, \href{http://arxiv.org/abs/1112.0871}{arXiv:1112.0871 [cond-mat.dis-nn]}.

\bibitem{Iguchi} K.~Iguchi, \textit{Theory of quasiperiodic lattices. II. Generic trace map and invariant surface}, \href{http://prb.aps.org/abstract/PRB/v43/i7/p5919_1}{Phys. Rev. \textbf{B 43}, 5919 (1991)}.



\end{thebibliography}
\end{document}